\documentstyle[12pt,epsfig]{article}

\newcommand{\be}{\begin{equation}}
\newcommand{\ee}{\end{equation}}
\newcommand{\bea}{\begin{eqnarray}}
\newcommand{\eea}{\end{eqnarray}}
\newcommand{\inli}{\int\limits}

\newcommand{\bkp}{{\bf k}_\perp}
\newcommand{\bqp}{{\bf q}_\perp}

\def\la{\mathrel{\mathpalette\fun <}}

\def\fun#1#2{\lower3.6pt\vbox{\baselineskip0pt\lineskip.9pt
\ialign{$\mathsurround=0pt#1\hfil##\hfil$\crcr#2\crcr\sim\crcr}}}

\begin{document}

\title{ $D^+_s \to \pi^+\pi^+\pi^-$ decay: the $1^3P_0s\bar s$
component in scalar-isoscalar mesons}

\author{V.V. Anisovich, L.G. Dakhno and V.A. Nikonov}

\date{February 17, 2003}
\maketitle

\begin{abstract}

On the basis of data on the decay $D^+_s \to \pi^+\pi^+\pi^-$,
which goes dominantly via the transition $D_s \to \pi^+ s\bar s$,
we evaluate the $1^3P_0s\bar s$ components in the
scalar-isoscalar resonances  $f_0(980)$,
$f_0(1300)$, $f_0(1500)$ and  broad state $f_0(1200-1600)$.
The data point to a large $s\bar s$ component in the
$f_0(980)$: $40\% \la s\bar s \la 70\% $.
Nearly $30\%$ of the $1^3P_0s\bar s$ component
flows to the mass region 1300--1500 MeV
being shared by $f_0(1300)$, $f_0(1500)$
and   broad state $f_0(1200-1600)$: the interference of these states
results in a peak near $1400$ MeV with the width around $200$ MeV.

\end{abstract}

\section{Introduction}

The meson yields in the decay $D^+_s \to \pi^+\pi^+\pi^-$ \cite{D+s}
evoked immediately a great interest and now they are actively discussed
(see \cite{klempt,van,mink, cheng} and references therein).
The matter is that in this decay the production
of strange quarkonium is dominant,
$D^+_s \to \pi^+s\bar s$,
with a subsequent transition $s\bar s \to f_J \to
\pi^+\pi^-$. Therefore, the reaction $D^+_s \to \pi^+\pi^+\pi^-$ may
serve us as a tool for the estimation of  $s\bar s$
components in the $f_J$ mesons. This possibility is particularly
important in context of the determination of  quark content of the
$f_0$ mesons, for  the classification of $q\bar q$ scalar-isoscalar
states is a key problem in the search for exotics.

In the $D_s^+$ decay the production of  $f_0$ mesons proceeds
mainly via spectator mechanism, Fig. 1a:  this very mechanism
implements  the transition $s\bar s \to f_0$. Besides, the spectator
mechanism provides a strong production of the $ \phi (1020)$ meson.
To evaluate the $s\bar s$ components in $f_0$ mesons,  we use
the process $D^+_s \to \pi^+ \phi (1020)$ as a standard: we
consider
the ratio $D^+_s \to \pi^+ f_0 /D^+_s \to \pi^+ \phi (1020)$,
where the uncertainty related to the coupling $c\to \pi^+ s$ is
absent.

Calculation of the transition of Fig. 1a is performed in the spectral
integration technique, this technique was intensively used for the
study of weak decays of $D$ and $B$ mesons, see \cite{Melikhov}
and references therein. The cuttings of the triangle
diagram related to the double spectral integrals are shown in
Fig. 1b.

In addition, the $\pi^+ f_0$ production can originate from the
$W$-annihilation process of
Figs. 1c,d. It is a relatively weak transition,
nevertheless we take it into account, and the reaction
$D^+_s \to \pi^+ \rho^0 $ serves us as a scale to determine
the $W$-annihilation coupling $c\bar s \to u\bar d$.
Let us emphasize that the processes shown in Fig. 1 are of the
leading order in terms of the $1/N_c$ expansion rule.

In Section 2 we present the data set used for the analysis and write
down the amplitudes for the spectator and $W$-annihilation
processes. The results of calculations are
presented and discussed in Section 3. In this Section we demonstrate
that (i) the $W$-annihilation contributes weakly to the
$f_J$-meson production, and (ii) the $1^3P_0$-state dominates  the
transition $s\bar s\to f_0$, while the production of the $2^3P_0s\bar
s$-component is relatively suppressed, so the transition $D^+_s \to
\pi^+s\bar s\to \pi^+f_0$ is in fact a measure for the $1^3P_0s\bar
s$-component in scalar--isoscalar mesons.

In Conclusion we sum up what the data on the decay
$D^+_s \to \pi^+\pi^+\pi^-$ tell us, in particular with respect to
the identification of the lightest scalar $q\bar q$ nonet.

\section{Data set and the amplitudes for the spectator and
 $W$-annihilation  mechanisms}

In this Section  we present the data  used in the analysis
and write formulae  for the
spectator and $W$-annihilation
mechanisms , Fig. 1a and Figs. 1c,d,
respectively.

\subsection{The data set}

 In the recently measured spectra from the reaction
$D^+_s \to \pi^+\pi^+\pi^-$ \cite{D+s}, the relative weight of
channels
$\pi^+f_0(980)$ and $\pi^+\rho^0(770)$
is evaluated,
\be
BR\left (\pi^+f_0(980)\right)= 57\%\pm 9\%, \qquad
BR\left(\pi^+\rho^0(770)\right)= 6\%\pm 6\%\ ,
\label{2.1.1}
\ee
and the ratio of yields,
\be
\Gamma (D^+_s \to \pi^+\pi^+\pi^-)
/\Gamma (D^+_s \to \pi^+\phi(1020))=
0.245\pm 0.028^{+0.019}_{-0.012}\,
\label{2.1.2}
\ee
is measured. These values are the basis to determine
relative weight of the $s\bar s $ component in the
$f_0(980)$.

Besides, in ref. \cite{D+s} the bump in the wave $(IJ^{PC}=00^{++})$
is seen at $1434\pm 18$ MeV
with the width $173\pm 32$ MeV; this should be a contribution from
the nearly located resonances $f_0(1300)$,  $f_0(1500)$
and the broad state $f_0(1200-1600)$.  Relative weight of this bump
is equal to:
\be
BR \left
(\pi^+  (f_0(1300)+f_0(1500)+f_0(1200-1600))\right )=26\%\pm 11\% \; .
\label{2.1.3}
\ee
This magnitude allows us to  determine the total weight of
the $1^3P_0s\bar s$ component in the states
$f_0(1300)$, $f_0(1500)$, $f_0(1200-1600)$.

Now the data of FOCUS
collaboration \cite{focus} on the decay
$D_s\to \pi^+f_0(980)$  are available.
 These data are compatible with
those of \cite{D+s}, so we do not use them in our estimates, and we
base on the ratios $\pi^+f_0/\pi^+\phi$ measured in \cite{D+s}.

In addition, the production of $f_2(1270)$ is seen in \cite{D+s}:
$BR\left (\pi^+f_2(1270)\right)= 20\%\pm 4\%$ that makes it necessary
to include tensor mesons into the calculation machinery.

\subsection{Decay amplitudes and partial widths}

The spin structure of the amplitude depends on  the type of the
produced meson --- it is different for scalar $(f_0)$, vector
$(\phi,\omega,\rho)$ or tensor $(f_2)$ mesons. Let us denote the
momenta of the produced scalar $(S)$, vector $(V)$ and tensor $(T)$
mesons by $p_M$ where $M=S,\, V,\, T$; the $D_s$-meson momentum is
referred as $p$.

The production amplitude is written as
\be
A(D_s\to \pi^+M)=\widehat O _M(p,p_M)A_M(q^2)\, ,
\label{2.2.1}
\ee
where the spin operators $\widehat O _M(p,p_M)$ for
scalar, vector and tensor mesons read as follows:
\be
\widehat O _S(p,p_S)=1\ ,\quad \widehat O _V(p,p_V)=p_{V\perp \mu}\ ,\quad
\widehat O _T(p,p_T)=
\frac{p_{T\perp \mu}p_{T\perp
\nu}}{p^2_{T\perp}} -\frac13g^\perp_{\mu\nu}  \, .
\label{2.2.2}
\ee
The momenta $p_{V\perp}$ and $p_{T\perp}$ are
orthogonal to the $D_s$-meson momentum $p$:
\be p_{V\perp
\mu}=g^\perp_{\mu\mu'}p_{V\mu'},\quad p_{T\perp
\mu}=g^\perp_{\mu\mu'}p_{T\mu'} ,\quad
g^\perp_{\mu\mu'}=g_{\mu\mu'}-\frac {p_\mu p_{\mu'}}{p^2}\ .
\label{2.2.3}
\ee
In the spectral integration technique, the invariant
production amplitude $A_M(q^2)$ is calculated as a function of
$q^2=(p-p_M)^2=m_{\pi}^2$.

In terms of the spin-dependent operators $\widehat O _M(p,p_M)$,
partial width for the transition $D_s^+\to \pi^+M$ reads:
\be
m_{D_s}\Gamma(D_s^+\to \pi^+ M)=\left |F_M(q^2=m^2_\pi)\right |^2
\left ( \widehat O _M(p,p_M) \right )^2
\frac{\sqrt{-p^2_{M\perp}}}{8\pi m_{D_s}}\ ,
\label{2.2.4}
\ee
where
\be
\left (\widehat O _S(p,p_S)\right )^2=1\ ,
\quad \left (\widehat O
_V(p,p_V)\right )^2=-p^2_{V\perp \mu}\ ,
\quad \left (\widehat O _T(p,p_T)
\right )=
\frac{2}{3}\ .
\label{2.2.5}
\ee
In (\ref{2.2.4}), the value $\sqrt{-p^2_{M\perp}}$ is equal
to the center-of-mass relative
momentum of mesons in the final state; it is determined by the
magnitudes of the meson masses as follows:
$$\sqrt{-p_{M\perp}^2}=
\sqrt{[m^2_{D_s}-(m_M+m_\pi)^2][m^2_{D_s}-(m_M-m_\pi)^2]}
/2m_{D_s}\ .
$$

\subsection{Amplitudes for the spectator
and $W$-annihilation processes in the light-cone variables}

In the leading order of the $1/N_c$ expansion,
there exist two types of processes which govern the decays
$D^+_s\to \pi^+f_0$, $\pi^+f_2$, $\pi^+\phi/\omega$,
$\pi^+\rho^0$. They are shown in Figs. 1a,c,d. We refer to the
process of Fig. 1a as a spectator  one, while that shown in Fig. 1c,d
is called the $W$-annihilation  process.
The transition $s\bar s \to meson$ is a characteristic feature of the
spectator mechanism, it contributes to the production of
isoscalar mesons: $D^+_s\to \pi^+f_0,\, \pi^+f_2,\, \pi^+\phi,\, \pi^+
\omega$, whereas  $\rho^0$ cannot be produced within  spectator
mechanism. The $W$-annihilation  contributes  to the production of
mesons with both $I=1$ and $I=0$,
 $D^+_s\to \pi^+f_0,\, \pi^+f_2,\, \pi^+\phi,\, \pi^+ \omega$
and $D_s\to \pi^+\rho^0$. Therefore, the latter reaction, $D_s\to
\pi^+\rho^0$, allows us to evaluate relative weight of the
effective coupling constant for $W$-annihilation, thus giving a
possibility to estimate the $W$-annihilation contribution to the
channels of interest: $D^+_s\to \pi^+f_0$, $\pi^+f_2$, $\pi^+\phi$.
This estimate tells us that the $W$-annihilation is
relatively weak that agrees with conventional evaluations,
see, for example, \cite{Melikhov}.

The amplitudes for the spectator production of mesons
(Fig. 1a) and for $W$-annihilation
 (Figs. 1c,d) can be calculated in terms of double
spectral integral representation developed for the quark
three-point diagrams in
\cite{Melikhov,pi_eta}. The calculation scheme for the
diagram of Fig. 1a in the spectral integration technique
is as follows.
We consider the relevant energy-off-shell diagram shown in Fig. 1b for
which the momentum of the $c\bar s$ system, $P=k_1+k_2$, obeys the
requirement $P^2\equiv s> (m_c+m_s)^2$, while the $s\bar s$ system,
with the momentum $P'=k'_1+k_2$, satisfies the constraint
$P'^2\equiv s'> 4m_s^2$, here $m_{s,c}$ are the masses of the
constituent $s,c$ quarks which are taken to be $m_s=500$ MeV and
$m_c=1500$ MeV. The next step consists in the calculation of double
discontinuity of the triangle diagram (cuttings I and II in Fig. 1b)
which correspond to real processes, and the double discontinuity is the
integrand of  double dispersion representation.

The double dispersion integrals may be rewritten in terms of the
light-cone variables, by introducing the light-cone wave functions
for the $D^+_s$-meson and produced mesons $ f_0,\, f_2,\, \phi,\,
\omega, \, \rho^0$: the calculations performed here are done by using
these variables.

Our calculations have been carried out in the limit of negligibly small
pion mass, $m_\pi \to 0$, that is a reasonable approach, for in
the ratio $D_s\to \pi^+ f_J/D_s \to \pi^+ \phi$ the uncertainties
related to this limit are mainly cancelled.

\subsubsection{Spectator-production form factor}

The form factor for
the spectator process given by the triangle diagram of Fig. 1a reads:
\be
F^{(spectator)}_M (q^2)= \frac{G_{spectator}}{16\pi^3}\inli_0^1
\frac{dx}{x(1-x)^2} \int d^2k_\perp \psi_{D_s}(s)\psi_M(s')
S_{D_s\to \pi M}(s,s',q^2)\ .
\label{2.3.1}
\ee
Here $G_{spectator}$ is the vertex for the decay transition $c\to \pi^+
s$; the light-cone variables $x$ and $\bkp$ refer to the momenta of
quarks in the intermediate states.
The energies squared for initial  and
final quark states ($c\bar s$ and $s\bar s$)
are written in terms of the light-cone variables as follows:
\be
s=\frac{m^2_c+k^2_\perp}{1-x}+\frac {m^2_s+k^2_\perp}{x}\ ,\quad
s'=\frac{m^2_s+ (\bkp +x\bqp )^2 }{x(1-x)}\ .
\label{2.3.2}
\ee
The limit of
the negligibly small pion mass corresponds to $\bqp \to 0$; in
(\ref{2.3.1}) this limit is attained within
numerical  calculation of $F^{(spectator)}_M (q^2)$ at small negative
$q^2$.

In the spectral integration technique, wave functions of the initial
and final states are determined as ratios of vertices to the
dispersion-relation denominators,
$\psi_{D_s\to c\bar s}(s)=G_{D_s\to c\bar s}(s)/(s-m^2_{D_s})$ and
$\psi_{M\to s\bar s}(s')=G_{M\to s\bar s}(s')/(s'-m^2_M)$, see
\cite{Melikhov,pi_eta} for details.

In our calculations, the  $D^+_s$-meson wave functions and
$s\bar s$ component are pa\-ra\-met\-ri\-zed as follows:
\be
\psi_{D_s=c\bar s}(s) = C_D\exp (-b_Ds)\ , \qquad
\psi_M(s')= C_M\exp (-b_Ms')\ ,
\label{2.3.3}
\ee
where $C_D$ and $C_M$  are the normalization constants for the wave
functions and $b_D$ and $b_M$ characterize the mean radii squared of
the $c\bar s$ and $s\bar s $ systems, $R^2_{D_s}$ and $R^2_M$.  In the
approximation (\ref{2.3.3}), the mean radii squared are the only
parameters for the description of quark wave functions. Based
on the results of the analysis of the radiative decays \cite{phi} we
put $R^2_M$ for $ f_0(980)$, $\phi (1020)$ and $f_2(1270)$ to
be of the order of the pion
radius squared, $R^2_M\sim R^2_\pi =10$ GeV$^{-2}$, that corresponds to
the following wave function parameters  for $s\bar s$
components (in GeV units):
\be
b_{f_0(980)}=1.25, \quad  b_{\phi (1020)}=2.50, \quad
b_{f_2(1270)}=1.25 \, ,
\label{2.3.4}
\ee
$$
C_{f_0(980)}=98.92, \quad  C_{\phi (1020)}=374.76, \quad
C_{f_2(1270)}=68.85 \, .
$$
Recall that in \cite{phi} the mean radius squared was defined
through the $Q^2$-dependence of meson form factor at small momentum
transfers, $F_M(Q^2)\simeq 1-Q^2 R_M^2/6$, and
the normalization factor
$C_M$ is given by $F_M(0)=1$, that actually represents the convolution
$\psi_M \otimes \psi_M =1$.

For the $D^+_s$-meson, the charge radius squared
$R^2_{D_s}$ is of the order of $3.5\div 5.5$ GeV$^{-2}$
\cite{Melikhov} that corresponds to
\be
b_{D_s}\simeq (0.70 \div  1.50 )\, \rm{GeV^{-2}} \ ,\qquad
C_{D_s}\simeq (157.6  \div   7205.4 )\, \rm{GeV^{-2}} .
\label{2.3.5}
\ee
One should keep in mind that the $D^+_s$-meson charge radius squared
is determined by two form factors,
$F_c(Q^2)$ and $F_{\bar s}(Q^2)$,
when the photon interacts with $c$ and $\bar s$ quarks:
$2 F_c(Q^2)/3+ F_{\bar s}(Q^2)/3 \simeq 1- R^2_{D_s}Q^2/6$.

The factor $S_{D_s\to \pi M}(s,s',q^2)$ is defined by the  spin
structure of the quark loop in the diagram of Fig. 1b.
Corresponding trace of the three-point quark loop
is equal to:
\be
\widehat S_M = -{\rm Tr}\, \left [\widehat Q_M(-\hat k_2+m_s)i\gamma_5
(\hat k_1+m_c)i\gamma_5(\hat k'_1+m_s) \right ]\ ,
\label{2.3.6}
\ee
where $i\gamma_5$ stands for the $D_s$-meson and pion vertices, and
$\widehat Q_M$ is the spin operator for the transition
$(s\bar s \to meson\, M)$
which is defined  for scalar, vector and tensor mesons as follows:
\bea
\widehat Q_S&=&1\ , \nonumber \\
\widehat Q_V&=&\gamma'_{\perp\mu}\ , \nonumber \\
\frac12\widehat Q_T&=&k'_\mu\gamma'_{\perp\nu}+
k'_\nu\gamma'_{\perp\mu}-
\frac23\hat k'g'^\perp_{\mu\nu}\ .
\label{2.3.7}
\eea
Here $k'=(k'_1-k_2)/2$ and  $\gamma'_{\perp\mu}=g'^\perp_{\mu\nu}
\gamma_{\perp\nu}$, where $g'^\perp_{\mu\nu}=g_{\mu\nu}-P'_{\mu}
P'_{\mu'}/P'^2$. The operator $\widehat Q_T$ stands for the production
of $f_2$-mesons belonging to the basic $1^3P_2 q\bar q$ multiplet.

With these definitions, the factor $\widehat S_{D_s\to \pi M}
(s,s',q^2)$ can be
calculated through normalized convolution of the quark-loop operator
(\ref{2.3.6}) with the spin operator of the amplitude given by
(\ref{2.2.2}) but  determined in the space of internal momenta,
by substituting    $p\to P$ and $p_{M}\to P'$, namely,
\be
S_{D_s\to \pi M}(s,s',q^2)=
\frac{\left (\widehat S_M\cdot \widehat O_M(P,P')\right )}
{\left (\widehat O_M(P,P')\right )^2}\ .
\label{2.3.8}
\ee
Let us emphasize once again, that we calculate the integrand of the
spectral integral for the energy-off-shell process of Fig. 1b. Because
of that, the invariant spin-dependent structure
$S_{D_s\to \pi M}(s,s',q^2)$ should be calculated through
(\ref{2.3.8}) with the energy-off-shell operators $\widehat O_M(P,P')$
and mass-on-shell constituents. The spin factors
determined by (\ref{2.3.8}) read:
\bea
S_{D_s\to \pi f_0}(s,s',q^2)&=&
2(s m_s - 2 m_c^2 m_s - m_c s' + 4 m_c m_s^2 + q^2 m_s - 2 m_s^3)\ ,
\nonumber \\
S_{D_s\to \pi \phi/\omega}(s,s',q^2)&=&
\frac{-8 s'}\lambda
(s m_c^2-s m_c m_s-s q^2-m_c^4+2 m_c^3 m_s-m_c^2 s'
\nonumber \\
&+&m_c^2 q^2 +m_c s' m_s-m_c q^2 m_s-2 m_c m_s^3+m_s^4)\ ,
\nonumber \\
S_{D_s\to \pi f_2}(s,s',q^2)&=&
\frac{8 s'}\lambda
(s m_c^2-s m_c m_s-s q^2-m_c^4+2 m_c^3 m_s-m_c^2 s'
\nonumber \\
&+&m_c^2 q^2+m_c s' m_s-m_c q^2 m_s-2 m_c m_s^3+m_s^4)
\nonumber \\
&\times& (s-2 m_c^2-s'+q^2+2 m_s^2)\ ,
\label{2.3.9}
\eea
where $\lambda=(s'-s)^2-2q^2(s'+s)+q^4$. In the performed calculations
we have used the moment-expansion technique, for the details see
the review paper \cite{operator} and references therein.

\subsubsection{$W$-annihilation form factor}

The right-hand side three-point block of Fig. 1c,d which describes the
transitions of the $u\bar d$ system into mesons
$\pi^+ M$ can be also
calculated with the formulae similar to (\ref{2.3.1}). One has:
\be
F^{(W)}_M (q^2)=
\frac1{16\pi^3}\inli_0^1 \frac{dx}{x(1-x)^2} \int d^2k_\perp
\frac{G_w}{s-m^2_{D_s}-i0}\psi_M(s')
S_{D_s(u\bar d)\to \pi M}(s,s',q^2)\ .
\label{2.3.10}
\ee
For the transition $D_s^+\to u\bar d$ we use the
dispersion relation description, and vertex $G_W$ is treated as
energy-independent factor.

The spin factor $S_{D_s(u\bar d)\to \pi M}(s,s',q^2)$
is determined by the
triangle graph of Fig. 1c,d, with light quarks in the intermediate
state. Therefore,
\be
\widehat S_M^{(W)}= -{\rm Tr}\,
\left [\widehat Q_M(-\hat k_2+m)i\gamma_5
(\hat k_1+m)i\gamma_5(\hat k'_1+m) \right ]\ ,
\label{2.3.11}
\ee
where $m=350$ MeV, and $\widehat Q_M$ is given by (\ref{2.3.7}).
Furthermore, the spin factors
$S_{D_s(u\bar d)\to \pi M}(s,s',q^2)$ is
calculated with the use of
(\ref{2.3.8}).  For scalar, vector and tensor mesons, respectively,
they are equal to
\bea
S_{D_s(u\bar d)\to \pi
f_0}(s,s',q^2)&=&2m(s-s'+q^2)\ ,\nonumber \\ S_{D_s(u\bar d)\to
\pi\phi/\omega }(s,s',q^2)&=&\frac{8ss'q^2}{\lambda}\ , \nonumber \\
S_{D_s(u\bar d)\to \pi f_2}(s,s',q^2)&= &
-\frac{8ss'q^2}{\lambda}(s-s'+q^2)\ .
\eea
The transition amplitude defined by Eq. (\ref{2.3.10}) is
complex-valued.

\section{Calculations and results}

Here we write down the amplitudes used for the calculation of the decay
processes --- corresponding results are presented below.

\subsection{ Amplitudes for decay channels
$\pi^+f_0(980)$, $\pi^+\phi(1020)$,\\
$\pi^+f_2(1270)$, $\pi^+\rho^0$}

Taking into account
two decay processes, spectator and $W$-annihilation, we write the
transition amplitude as follows:
\be
A(D_s\to \pi^+M)=\xi^{(spectator)}_M
F^{(spectator)}_M(0)+ \xi^{(W)}_M F^{(W)}_M(0)\ ,
\label{3.1}
\ee
where the factors $\xi^{(spectator)}$ and $\xi^{(W)}$
are determined by flavour content of isoscalar mesons.
In terms of the quarkonium states
$s\bar s$ and $n\bar n=(u\bar u+d\bar d)/\sqrt{2}$,
we define  flavour wave functions of  isoscalar mesons as
\bea
\phi(1020)&:&\qquad n\bar n \sin\varphi_V+ s\bar s \cos\varphi_V\ ,
\nonumber \\
f_0(980)&:&\qquad n\bar n \cos\varphi [f_0(980)]+
s\bar s \sin\varphi [f_0(980)]\ ,
\nonumber \\
f_2(1270)&:&\qquad n\bar n \cos\varphi_T+ s\bar s \sin\varphi_T\ .
\label{3.2}
\eea
that serves us for the determination of  coefficients in
(\ref{3.1}):
\bea
D_s^+&\to &\pi^+\phi(1020):\quad
\xi^{spectator}_\phi=\cos \varphi_V\ ,\quad
\xi^{W}_\phi=\sqrt{2}\sin \varphi_V\ ,
\nonumber \\
D_s^+&\to &\pi^+f_0(980):\quad
\xi^{spectator}_{f_0(980)}=\sin \varphi [f_0(980)] , \quad
\xi^{W}_{f_0(980)}=\sqrt{2}\cos \varphi [f_0(980)]\ ,
\nonumber \\
D_s^+&\to &\pi^+f_2(1270):\quad
\xi^{spectator}_{f_2(1270)}=\sin \varphi_T\ , \quad
\xi^{W}_{f_2(1270)}=\sqrt{2}\cos \varphi_T\ .
\label{3.3}
\eea
For $\phi (1020)$, which is dominantly the $s\bar s$
state, we fix  mixing angle in the interval
$|\varphi_V|\le 4^\circ$.

The production of $\pi^+\rho^0$ is due to the direct
mechanism only:
\be
D^+_s\to \pi^+\rho^0:\qquad \xi^{(spectator)}_\rho=0\ , \qquad
\xi^{(W)}_\rho=\sqrt{2}\ .
\label{3.4}
\ee

\subsection{Evaluation of the ratio $G_{W}/G_{spectator}$}

To evaluate the ratio $G_{W}/G_{spectator}$
we use the reaction $D_s^+ \to \pi^+ \rho^0$,
with the experimental constraint
$\Gamma(\pi^+\rho^0)/\Gamma(\pi^+\phi)\le 0.032$.
By using maximal value of $\Gamma(\pi^+\rho^0)/\Gamma(\pi^+\phi)=0.032$
we get the following ratios $F_M^{(W)}(0)/F_M^{(spectator)}(0)$ for
scalar, vector and tensor mesons at $R^2_{D_s}=4.5$ GeV$^{-2}$:
\bea
\frac{\sqrt{2}
F_S^{(W)}(0)}{F_S^{(spectator)}}&& =(0.28+i0.75)\cdot 10^{-3}\ ,
\nonumber \\
\frac{\sqrt{2}
F_V^{(W)}(0)}{F_V^{(spectator)}}&& =(0.5+i15.1)\cdot 10^{-2} \ ,
\nonumber \\
\frac{\sqrt{2}
F_T^{(W)}(0)}{F_T^{(spectator)}}&& =(0.26+i1.35)\cdot 10^{-3} \ .
\eea
This evaluation tells us that the $W$-annihilation
contribution is comparatively small, and one may neglect it when the
reactions $D^+_s\to \pi^+ f_0$ and $D^+_s\to \pi^+ f_2$ are studied.

\subsection{Evaluation of relative weights of the
$1^3P_0 s\bar s$ and $2^3P_0 s\bar s$ states for the decay
$D^+_s\to \pi^+ f_0$}

In the region 1000--1500 MeV one can expect the existence of
scalar-isoscalar states which belong to the basic and first
radial-excitation
$q\bar q$-nonets, $1^3P_0 $ and $2^3P_0 $. Here we estimate
relative weights of the states
$1^3P_0 s\bar s$ and $2^3P_0 s\bar s$ in the  transitions
$D^+_s\to \pi^+ \, s\bar s \to \pi^+ f_0$.

The form factor for the
production of  radial-excitation state is given by
(\ref{2.3.1}), with a choice of the wave function as follows:
\be
\psi_{M(rad.\,excit.)}(s')= C_{rad.\,excit.}
(d_{rad.\,excit.}s'-1)\exp (-b_{rad.\,excit.}s')\ ,
\label{3.6}
\ee
Two parameters in (\ref{3.6}) can be determined by the normalization
and orthogonality conditions,
$\psi_{M(rad.\,excit.)}\otimes \psi_{M(rad.\,excit.)}=1$
and $\psi_{M(rad.\,excit.)}\otimes \psi_{M(basic)}=0$, while the
third one
can be related to the mean radius squared, $R_{rad.\,excit.}^2$. In our
estimates we keep $R_{rad.\,excit.}^2$ to be of the order of pion
radius squared, or larger,
$1\le R_{rad.\,excit.}^2/R_{\pi}^2 \le 1.5$. To be
precise, we present as an example the wave function parameters
(in GeV units) for $2^3P_0 s\bar s$
state with $R_{rad.\,excit.}^2 =11.3$
GeV$^{-2}$:
\be
b_{rad.\,excit.}=1.75, \quad C_{rad.\,excit.}=938.5,
\quad d_{rad.\,excit.}=0.60.
\label{3.7}
\ee
In Fig. 2 we show the ratios
$$
\frac{\Gamma \left (D^+_s \to \pi^+(2^3P_0s\bar s) \right )}
{\Gamma \left (D^+_s \to \pi^+(1^3P_0s\bar s) \right )}
$$
for different values of radius square of the
$(2^3P_0s\bar s)$  component, $R^2(f_0^{rad.\,excit.})$,
in the interval $3.5 \la R^2_{D_s}\la 5.5$
GeV$^{-2}$.

From the point of view of the calculation technique, a suppression of
the production of the
$2^3P_0\,s\bar s$ state in the process of Fig. 1a is due to the
existence of zero in the wave function $\psi_{M(rad.\,excit.)}$,
see (\ref{3.6}).
As a result, the convolution of wave functions
$\psi_{D_s}\otimes \psi_{M(rad.\,excit.)}$ occurred to be
considerably less than the convolution $\psi_{D_s}\otimes \psi_M$.

So, the production of the
radial-excitation state $D^+_s\to \pi^+(2^3P_0s\bar s)$ is relativelly
suppressed, by the factor of the order of $1/30$. This means
that by measuring $f_0$-resonances one measures actually the yield
of the $1^3P_0\,s\bar s$ state.

\subsection{The decay $D^+_s\to \pi^+ f_0(980)$}

The channel $D^+_s\to \pi^+ f_0(980)$ dominates the decay
$D^+_s\to \pi^+\pi^+\pi^-$, it comprises $57\% \pm 9\%$. Taking into
account the branching ratio $BR\left (f_0(980)\to
\pi^+\pi^-\right)\simeq 53\%$
\cite{K} and the ratio of yields (\ref{2.1.2}), we have:
\be
\frac{\Gamma\left (D^+_s\to \pi^+ f_0(980) \right)}
{\Gamma\left (D^+_s\to \pi^+ \phi(1020) \right)}=0.275(1\pm 0.25)\ .
\label{3.9}
\ee
Calculations performed with formulae (\ref{2.3.1}), (\ref{2.3.10})
and (\ref{3.1}) allow one to fulfill this ratio with
\be
35^\circ\le
|\varphi|\le 55^\circ\ ,
\label{3.10}
\ee
keeping the charge radius of the
$D^+_s$-meson  in the interval $3.5 \;{\rm GeV}^{-2} \le R^2_{D_s}
\le 5.5 \;{\rm GeV}^{-2}$, see Fig. 3a.

The data on radiative decays $f_0(980)\to \gamma\gamma$ and
$\phi(1020)\to \gamma f_0(980)$ tell us that the mixing angle $\varphi$
can be either $\varphi=-48^\circ\pm 6^\circ$ or $\varphi=83^\circ\pm
4^\circ$
\cite{phi}. The constraint (\ref{3.10}) shows that      the
$D^+_s$-meson decay prefers the
solution with negative mixing angle, thus supporting the $f_0(980)$ to
be dominantly flavour-octet state.
The analysis of  hadronic spectra
in terms of the K-matrix approach \cite{K,km1500} also points
to the flavour-octet origin of the $f_0(980)$.

\subsection{The decay $D_s^+ \to \pi^+f_2(1270)$}

Taking into account $BR[f_2(1270)\to \pi^+\pi^-]\simeq 57\% $,
one has
\be
\frac{\pi^+f_2(1270)}{\pi^+\phi(1020)}=0.09(1\pm 0.2)\ .
\label{3.11}
\ee
The values of $\varphi_T$, which satisfy the ratio (\ref{3.11}),
are shown in Fig. 3b.
With $ R^2_{D_s}\simeq 3.5 \div 5.5\; {\rm GeV}^{-2}$
we have $|\varphi_T |\simeq 20^\circ - 40^\circ$, that does not
contadict the data both on  hadronic decays and the radiative decay
$f_2(1270)\to \gamma\gamma$ which give $\varphi_T \simeq 0 - 20^\circ$.
Therefore, the production of $f_2(1270)$ in $D_s^+ \to \pi^+\pi^+\pi^-$
agrees reasonably with the weight of  $s\bar s$ component measured
in other reactions.

\subsection{Bump near $1430$ MeV: \\ the decays $D_s^+ \to \pi^0
\left ( f_0(1300)+f_0(1500)+ f_0(1200-1600)\right )$}

In the region 1300--1500 MeV two comparatively  narrow resonances
$f_0(1300)$, $f_0(1500)$ and
the broad state $ f_0(1200-1600)$ are located (for more detail see
\cite{K,ufn02} and references therein). These resonances
are the mixtures of quarkonium states from the multiplets
$1^3P_0 $ and $2^3P_0 $ and scalar
gluonium $gg$:
\be
1^3P_0 s\bar s, \quad 1^3P_0 n\bar n, \quad
2^3P_0 s\bar s, \quad 2^3P_0 n\bar n, \quad  gg.
\label{3.12}
\ee
We denote the probabilities for the $f_0$ resonance to have
$1^3P_0 s\bar
s$ and $ 2^3P_0 s\bar s$ components as $\sin ^2\varphi [f_0]$ and
$\sin ^2\varphi_{rad.\,excit.} [f_0]$, respectively. Then the amplitude
for the production of the $S$-wave $\pi^+\pi^-$ state due to decays
of $f_0(1300)$, $f_0(1500)$ and $f_0(1200-1600)$ reads:
\be
A\left (D^+_s \to \pi^+ (\pi^+\pi^-[\sim 1430 {\rm MeV}])_S \right )=
\sum\limits_n
\frac{m_n\sqrt{\Gamma(D^+_s\to\pi^+f_0(n))}
\sqrt{\Gamma_n(f_0\to\pi^+\pi^-)}}{m^2_n-s-im_n\Gamma_n}\ .
\label{3.13}
\ee
Here we are summing over the resonances
$n=f_0(1300),f_0(1500), f_0(1200-1600)$, with the following parameters
\cite{K,ufn02} (in GeV units):
\bea
\label{3.14}
f_0(1300):&& \quad
m=1.300, \quad \Gamma/2 = 0.12\ ,
\nonumber \\
f_0(1500):&&\quad m=1.500, \quad \Gamma/2 = 0.06\ ,
\nonumber \\
f_0(1200-1600):&& \quad
m=1.420, \quad \Gamma/2 = 0.508\ .
\eea
$\Gamma \left (D^+_s\to \pi^+f_0(n)\right )$ is determined by
Eqs. (\ref{2.3.1}) and (\ref{3.1}).

The peak which is seen in the $\pi^+\pi^-$ $S$-wave  near 1430 MeV is
determined by both  $s\bar s$ components in $ f_0(1300)$
and $f_0(1500) $ and relative phases of the amplitudes
of $ f_0(1300)$ and $f_0(1500)$ which govern the interference
with the background given by $f_0(1200-1400)$.
According to the analysis of hadronic decays \cite{K,ufn02}, the mixing
angles of these states are in the intervals
\bea
-25^\circ \la & \varphi [f_0(1300)] &\la 15^\circ \ ,
\nonumber \\
-3^\circ  \la & \varphi [f_0(1500)] & \la 18^\circ \ ,
\nonumber \\
28^\circ  \la & \varphi [f_0(1200-1600)] & \la 38^\circ \ .
\label{1300.1}
\eea
Different variants of the calculation of the $\pi^+\pi^-$ spectra near
1400 MeV are shown in Fig. 4 for the values of mixing angles
in the intervals (\ref{1300.1}).
The following parameters were used in the calculation of the
transitions
$D^+_s\to\pi^+ 1^3P_0 s\bar s\to\pi^+(\pi^+\pi^-)_S$:
\bea
&(1)&\varphi [f_0(1300)]=-7^\circ\ ,\quad
\varphi [f_0(1500)]= 7^\circ,\quad
\varphi [f_0(1200-1600)]= 37^\circ,
\nonumber \\
&(2)&\varphi [f_0(1300)]=-25^\circ,\quad
\varphi [f_0(1500)]= 17^\circ,\quad
\varphi [f_0(1200-1600)]= 37^\circ,
\nonumber \\
&(3)&\varphi [f_0(1300)]=\ 17^\circ,\quad
\varphi [f_0(1500)]= 17^\circ,\quad
\varphi [f_0(1200-1600)]= 37^\circ.
\eea
 The variants (1) and (2) in Fig. 4a reproduce well
the bump observed in the $D^+_s\to \pi^+\pi^+\pi^-$ decay: the relative
weight of the bump for variants (1), (2) is of the order of $\sim
15\%\div 20\%$ that agrees with measured weight of the peak
\cite{D+s}. The variant (3) demonstrates that with the same signs of
mixing angles for $f_0(1300)$ and $f_0(1500)$ the calculated curve
gives a dip, not  bump, near 1400 MeV.

In Fig. 4 one can also see the results of the calculation performed for
the radial excitation
state $2^3P_0 s\bar s$ (curves (4), (5)).
For transitions
$D^+_s\to\pi^+ 2^3P_0 s\bar s\to\pi^+(\pi^+\pi^-)_S$ the following
angles are used:
\bea
(4)&&\varphi_{rad.\,excit.}[f_0(1300)]=-7^\circ,\;
\varphi_{rad.\,excit.}[f_0(1500)]= 7^\circ,\;
\nonumber \\
&&\varphi_{rad.\,excit.}[f_0(1200-1600)]= 37^\circ,
\nonumber \\
(5)&&\varphi_{rad.\,excit.}[f_0(1300)]=-25^\circ,\;
\varphi_{rad.\,excit.}[f_0(1500)]= 17^\circ,\;
\nonumber \\
&&\varphi_{rad.\,excit.}[f_0(1200-1600)]= 37^\circ.
\eea
These curves
illustrate well the suppression rate of the production of the $2^3P_0
s\bar s$ in the decay $D^+_s\to \pi^+\pi^+\pi^-$.

The decay $D^+_s\to \pi^+\pi^+\pi^-$ was discussed in
\cite{van,mink,cheng} from the point of view of determination of the
quark structure of the resonances produced
--- let us stress the differences in the obtained results.
In the paper \cite{cheng} the hypothesis of the four-quark structure
of $f_0(980)$ is advocated: to explain a large yield of $f_0(980)$
within the four-quark model one needs to assume that the decay
$D^+_s\to \pi^+\pi^+\pi^-$ goes with a strong violation of the $1/N_c$
expansion rules. Recall that in terms of the $1/N_c$ expansion  the
processes shown in Fig. 1a,c,d dominate; it is not clear why these
rules, though working well in other decay processes, are violated in
$D^+_s\to \pi^+\pi^+\pi^-$. From the point of view of Ref.
\cite{cheng}, the yield of $s\bar s$ state is seen only at larger
masses, such as the 1400-MeV bump or higher.

In  papers \cite{van,mink}
the spectator mechanism shown in Fig. 1a is considered as a
dominant one:
the authors conclude that $s\bar s$ component dominates the resonance
$f_0(980)$ --- here our conclusions     are similar. However,
the interpretation of the bump around 1430 MeV  differs from
ours. Our calculations show that the production of $2^3P_0
s\bar s$ component is suppressed, so the bump at 1430 is a
manifestation of the $1^3P_0 s\bar s$ component, while it was stated
 in \cite{van,mink} that the component $2^3P_0 s\bar s$ is
responsible for this bump.

Let us emphasize again that, according to
our calculations, relative suppression of the
production of $2^3P_0 s\bar s$ component is due to the use of realistic
wave functions of radial-excitation states, see (\ref{3.6}): the
existence of a zero in the wave function results in a suppression of
the convolution $\psi_{D_s}\otimes \psi_{2^3P_0 s\bar s}$.

\section{Conclusion}

The performed calculations allow one to trace out the destiny of the
$s\bar s $ component in the dominating process $D^+_s \to \pi^+ +s\bar s
\to \pi^+ +f_0$ of Fig. 1a. We show that this transition is
realized mainly due to the $1^3P_0s\bar s$ state while the production
of radial-excitation state, $2^3P_0 s\bar s$, is suppressed by factor
$1/30$ or more. Therefore, the reaction $D^+_s \to \pi^+ +f_0$ is
a measure of the $1^3P_0s\bar s$ component in the $f_0$ mesons.

The data of Ref. \cite{D+s} tell us that $1^3P_0s\bar s$ is dispersed
as follows: about $2/3$ of this state is hold by the $f_0(980)$
and the last $1/3$ is shared between the states with masses in the
region 1300--1500 MeV which are $f_0(1300)$, $f_0(1500)$ and broad
state $f_0(1200-1600)$. This result is quite recognizable as concern
the percentage of the $f_0$-states produced in the decay
$D^+_s\to\pi^+ \pi^+ \pi^-$:
$BR\left(\pi^+f_0(980)\right)=57\%\pm9\%$ and
$BR\left(\pi^+f_0(bump\,at\,1430\,MeV)\right)=26\%\pm 11\%$.
The performed calculations  demonstrate that
nothing prevents this straightforward interpretation. In other words,\\
(i) the spectator mechanism $D^+_s \to \pi^+ +s\bar s \to \pi^+
+f_0$ dominates, \\
(ii) the production of the $2^3P_0$ states is
suppressed, and \\
(iii) the interference of the states $f_0(1300)$,
$f_0(1500)$, $f_0(1200-1600)$ may organize a bump with the mass $\sim
1400$ MeV and width $\sim 200$ MeV.

The data \cite{D+s} cannot provide us with more scrupulous information
about the weight of $1^3P_0s\bar s$ component in the resonances
$f_0(1300)$, $f_0(1500)$ and $f_0(1750)$ and the broad state
$f_0(1400-1600)$. To get such information, one needs to carry out a
combined analysis of the decay $D^+_s \to \pi^+ \pi^+ \pi^-$
and hadron reactions  with the
production of investigated resonances. The performed investigation
of the reaction $D^+_s \to \pi^+ \pi^+ \pi^-$ is quite in line with the
$K$-matrix analysis of hadronic reactions \cite{K,km1500,ufn02} that
tells us that  in scalar--isoscalar sector the lowest $1^3P_0q\bar q$
state is the flavour octet while the flavour singlet is a heavier one.

The authors thank A.V. Anisovich and A.V. Sarantsev for useful discussions.
The paper is supported by RFFI grant N 01-02-17861.

\newpage

\begin{figure}
\centerline{\epsfig{file=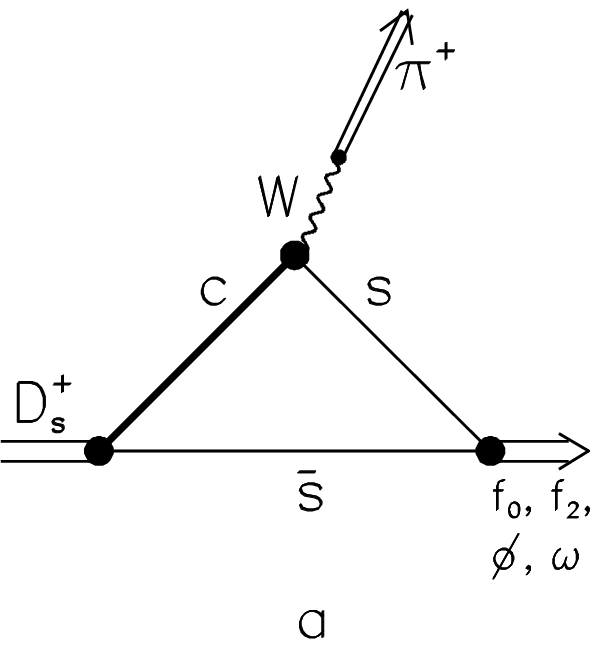,height=7.0cm}\hspace{0.5cm}
            \epsfig{file=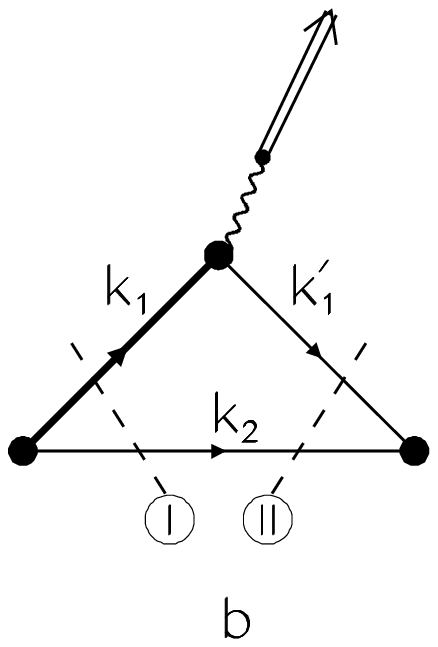,height=7.0cm}}
\vspace{1cm}
\centerline{\epsfig{file=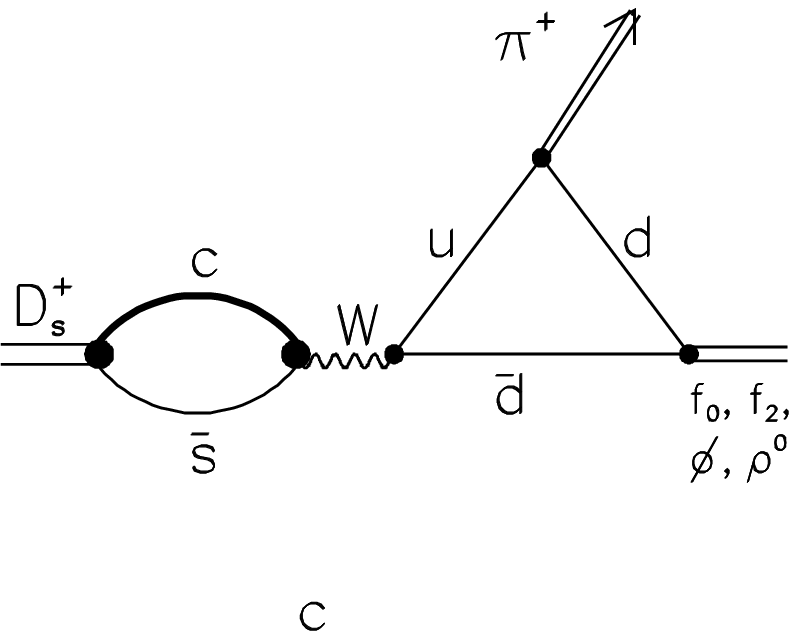,height=6.0cm}\hspace{0.9cm}
            \epsfig{file=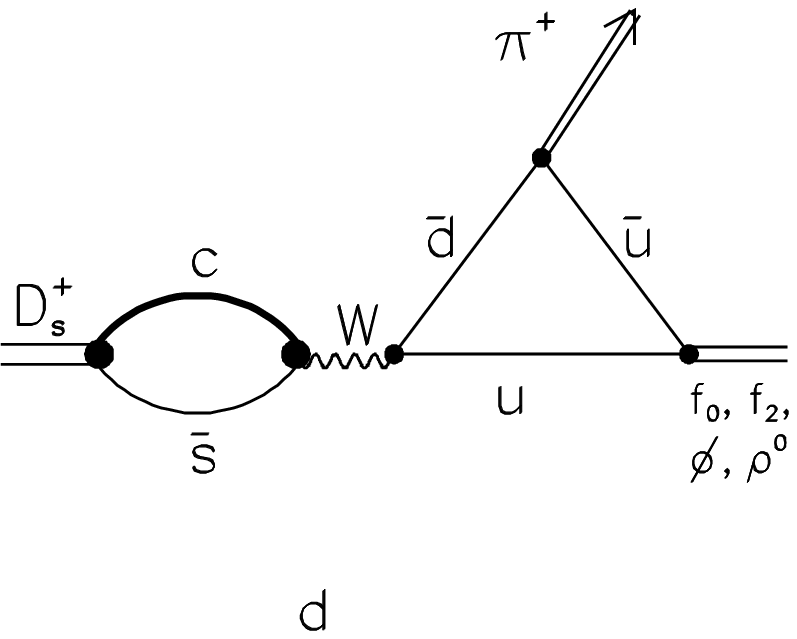,height=6.0cm}}
\caption{Diagrams determining the decay $D^+_s \to \pi^+\pi^+\pi^-$:
a) diagram for the spectator mechanism; b) energy-off-shell triangle
diagram for the integrand of the double spectral  representation;
 c,d) diagrams for the $W$-annihilation
mechanism  $D_s\to u\bar d$, with subsequent production of $u\bar u$
and $d\bar d$ pairs.}
\end{figure}

\begin{figure}
\centerline{\epsfig{file=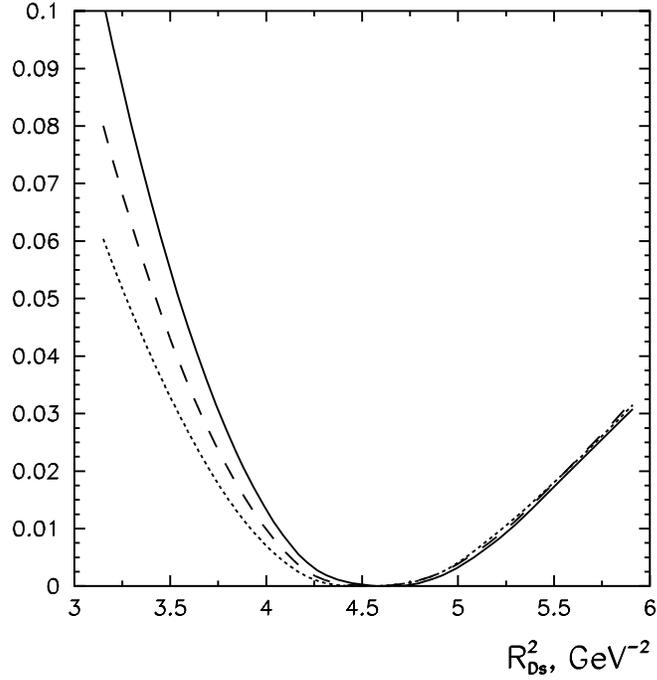,height=11.0cm}}
\caption{The ratio $\Gamma\left(D^+_s\to\pi^+ f^{rad.\,excit.}_0\right )/
\Gamma\left(D^+_s\to\pi^+ f^{basic}_0\right )$ as a function of
radius square of the $D^+_s$ meson, see Eq. (13). Calculations have
been carried out at fixed $R^2[f^{basic}_0]$=10 GeV$^{-2}$) for several
values of $R^2(f^{rad.\,excit.}_0)$: 10 GeV$^{-2}$ (solid line), 13
GeV$^{-2}$ (dashed line) and 16 GeV$^{-2}$ (dotted line).}
\end{figure}

\newpage

\begin{figure}
\centerline{\epsfig{file=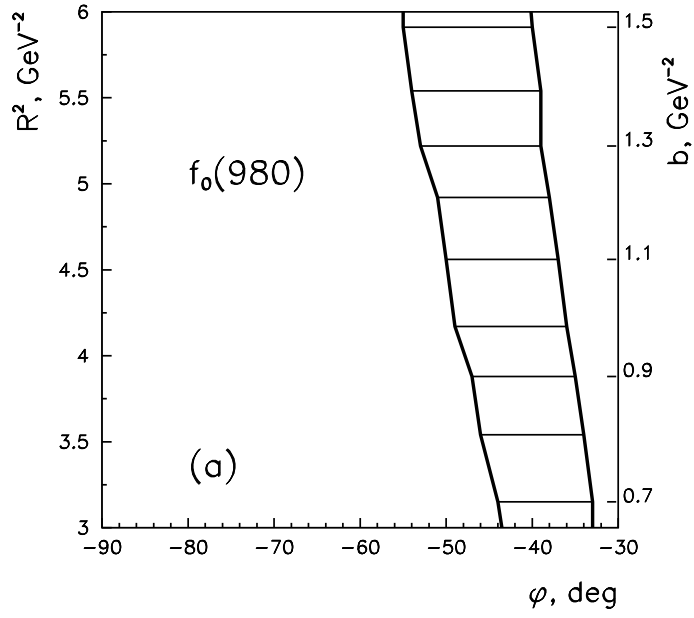,height=9.0cm}}
\centerline{\epsfig{file=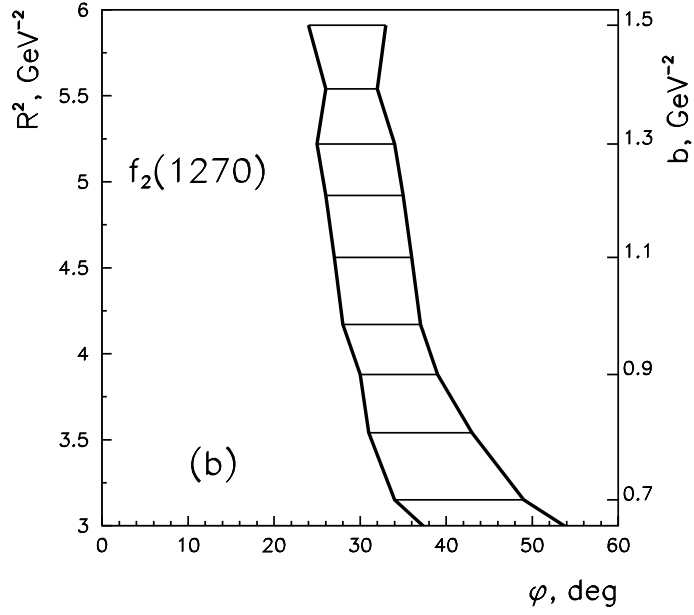,height=9.0cm}}
\caption{Allowed areas $(R^2_{D_s},\varphi_M)$
for flavour wave functions (22) provided by
experimental constraints  (\ref{3.9}) and (\ref{3.11}).}
\end{figure}

\newpage

\begin{figure}
\centerline{\epsfig{file=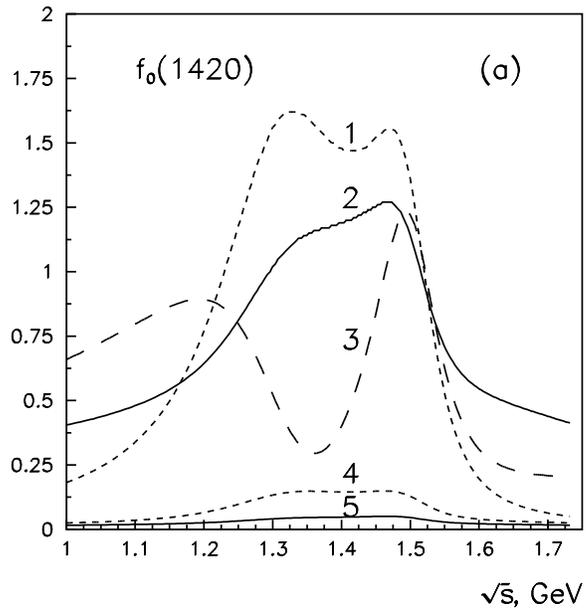,height=9.0cm}}
\centerline{\epsfig{file=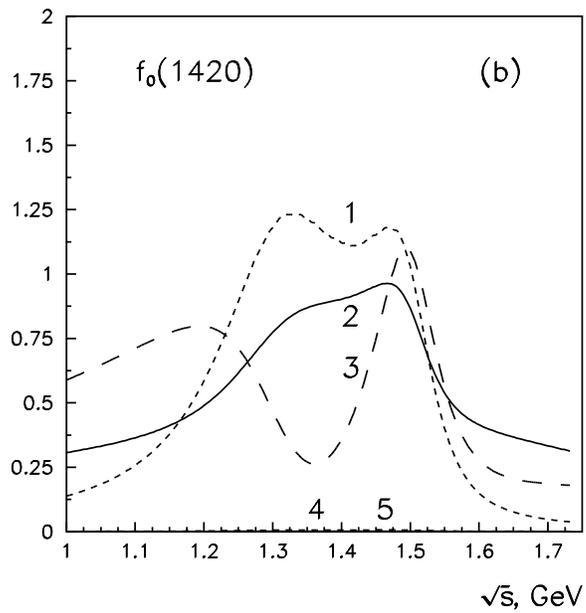,height=9.0cm}}
\caption{The $\pi^+\pi^-$  mass spectrum in the vicinity of 1430 MeV:
calculated curves respond to the  production of $f_0(1300)+f_0(1500)$
and the broad state $f_0(1200-1600)$, with
$R^2_{D_s}= 4.15$ GeV$^{-2}$ (a) and $R^2_{D_s}=5.90$ GeV$^{-2}$ (b).
Parameters used in calculations are given in (35) and (36).}
\end{figure}

\end{document}